\documentclass[aps,pra,preprintnumbers,nofootinbib,showpacs,twocolumn]{revtex4}

\usepackage{graphics}
\usepackage{graphicx}
\usepackage[colorlinks=true]{hyperref}
\usepackage{color}

\begin{document}

\title{Hall conductivity in the cosmic defect and dislocation space-time}

\author{Kai Ma
}
\email{makainca@yeah.net}

\author{Jian-hua Wang}

\affiliation{School of Physics Science, Shaanxi University of Technology, Hanzhong 723000, Shaanxi, P. R. China}

\author{Huan-Xiong Yang}
\affiliation{Interdisciplinary Center for Theoretical Study, University of Science and Technology of China, Hefei 200026, P. R. China}

\author{Hua-wei Fan}
\affiliation{School of Physics and Information Technology, Shaanxi Normal University, Xian 710000, Shaanxi, P. R. China}

\begin{abstract}
Influences of topological defect and dislocation on conductivity behavior of charge carries in external electromagnetic fields are studied. Particularly the quantum Hall effect is investigated in detail. It is found that the nontrivial deformations of spacetime due to topological defect and dislocation produce an electric current at the leading order of perturbation theory. This current then induces a deformation on the Hall conductivity. The corrections on the Hall conductivity depend on the external electric fields, the size of the sample and the momentum of the particle.
\end{abstract}



\maketitle

Topological defects, which encode the global geometry structure of the background space-time, can be realized in various physical systems ranging from the macroscopic scale to the microscopic scale, and therefore simulated extensive studies on its physical effects\cite{kibble, Vilenkin:string, Hiscock:1985, vilenkin:1985, Barriola:1989, Vilenkin:1983,kleinert, dzya,  katanaev, furtado, de-Assis:2000, furtado:landau-level, marques:landau-level, Bakke:landau-quantization, Bezerra, Bakke:2014-2, Bakke:2014-3, Bakke:2013-2, Bakke:phase-defect,  Bakke:2013-1, Bakke:2014-1, Montero:2011, Shen:2014, Ma:2013:TopoDefectSHE, Ma:2015:TopoDislocationSHE,Yang, Ma:SHE-NCS, Ma:SWmapNCSAC, Chowdhury:2014-2,Chowdhury:2013-3,Chowdhury:2013-1,Chowdhury:2013-4, Galtsov:1993, Lorenci:2002, Nakahara:1998, Hindmarsh:1995}. For instance, the cosmic string\cite{kibble, Vilenkin:string, Hiscock:1985, vilenkin:1985}, global monopole\cite{Barriola:1989}, and domain wall\cite{Vilenkin:1983, kleinert, dzya, katanaev, furtado, de-Assis:2000} can be formed at the phase transitions of the early universe from symmetric to asymmetric configurations. Even though topological defects have been commonly observed in condensed matter systems, so far there is no direct experimental evidence of the cosmological defects due to their extremely high energies. Therefore it is still attractive to explore the nontrivial geometry structures at the fundamental level. 

Various studies on the electromagnetic dynamics of the magnetic and electric dipole moments shown that both their global and local physical properties can be sensitive to non-trivial geometries\cite{Bakke:landau-quantization, Bezerra,Bakke:2014-2,Bakke:2014-3,Bakke:2013-2,Bakke:phase-defect, Bakke:2013-1,Bakke:2014-1,Montero:2011,Shen:2014,Ma:2013:TopoDefectSHE,Ma:2015:TopoDislocationSHE,Yang,Ma:SHE-NCS,Ma:SWmapNCSAC,Chowdhury:2014-2,Chowdhury:2013-3,Chowdhury:2013-1,Chowdhury:2013-4}. However, it is still difficult to observe them experimentally. On the other hand, it is naturally expected that the energy spectra of the charged carries in various quantum systems can be deformed by the topological defects, and hence can be used to detect the topological properties of the background spacetime. The effects of nontrivial geometries survive in the non-relativistic limit, and have been studied in Refs. \cite{Bakke:landau-quantization,Bakke:2014-2}. Nevertheless, the electromagnetic dynamics of charge carries in the presence of topological defects were less discussed. It was pointed out that the infinite degeneracy of the Landau levels is lifted by topological disclinations and defects\cite{furtado:landau-level,marques:landau-level}. In this letter we focus on conductivity behavior of the charger carries under the influences of topological defects and dislocation in the non-relativistic limit. We consider the quantum Hall effect which is one of the most profound phenomena of charge carries in electromagnetic fields. It is shown that the Hall conductivity receives a correction which depends linearly on the scale parameters of the topological defects, and are in the same order with the corrections on the spin-Hall conductivity\cite{Ma:2013:TopoDefectSHE,Ma:2015:TopoDislocationSHE}. However, we can expect that very steady and precise quantized Hall conductivity can provide better experimental sensitivity and can bound on parameters of the models.

We consider the quantum Hall effect in the presence of a topological defect. The dynamics of a relativistic Dirac particle in the curved spacetime is expressed in the generalized covariant form of the Dirac equation \cite{Nakahara:1998},
\begin{equation}\label{C-Dirac}
    [\tilde{\gamma}^{\mu}(x)(p_{\mu}- q A_{\mu}(x)-\Gamma_{\mu}(x))+mc^2]\psi(x)=0,
\end{equation}
$\Gamma_{\mu}(x)$ is the spinor connection, and $\tilde{\gamma}^{\mu}(x)$ are the elements of coordinate dependent Clifford algebra in the curved spacetime and satisfy the relation $\{\tilde{\gamma}^{\mu}(x),\tilde{\gamma}^{\nu}(x)\}=2g^{\mu\nu}(x)$, with $g^{\mu\nu}(x)$ being the matric of the spacetime in the presence of a topological defect. In the formalism of vierbein (or tetrad), which allows us to define the spinors in the curved spacetime, the metric has the form \cite{Nakahara:1998}, $g_{\mu\nu}(x)=e^{a}_{~\mu}(x)e^{b}_{~\nu}(x)\eta_{ab}$ and the coordinate dependent Dirac matrix reads $\tilde{\gamma}^{\mu}(x)=\gamma_{a}e^{a}_{~\mu}(x)$. The inverse of vierbein is defined by the relations, $e^{a}_{~\mu}(x)e^{\mu}_{~b}(x)=\delta^{a}_{~b}$ and $e^{\mu}_{~a}(x)e^{a}_{~\nu}(x)=\delta^{\mu}_{~\nu}$. 

The non-relativistic electromagnetic dynamics for a charged Dirac particle in the presence of topological defects has been discussed in Ref.\cite{Ma:2013:TopoDefectSHE} by using the Foldy-Wouthuysen transformation\cite{Foldy:nonrelativisitics}. It was shown that at the non-relativistic limit the Sch\"{o}rdinger equation becomes ( here we have neglected all the terms involving spin)
\begin{equation}\label{minic-ps-hamil}
     H
    =\frac{1}{2m}\bigg(\vec{p}- q\vec{A}(\vec{x})- q\vec{K}(\vec{p},\vec{x})\bigg)^2 +qV(\vec{x})~,
\end{equation}
where $\vec{K}$ behaving like an effective vector potential is the correction to the ordinary minimal coupling, but it is as functions of both position and momentum. In the convention of this work it is 
\begin{equation}\label{effectiveA-1}
\vec{K} = \frac{1}{q} \vec{\vec{\Omega}}\cdot \bigg( \vec{p} - q \vec{A}\bigg)~,
\end{equation}
where the matrix $\Omega$ is the correction to the ordinary verbena,
\begin{equation}\label{re-matric}
    \Omega^{a}_{~\mu}(x)=e^{a}_{~\mu}(x)-\delta^{a}_{~\mu},~~
    \Omega^{\mu}_{~a}(x)=e^{\mu}_{~a}(x)-\delta^{\mu}_{~a}.
\end{equation}

Next, we focus on the influences of the topological defect with line element given by
\begin{equation}\label{line-element-1}
     d s^{2}
    =c^2 d t^{2}- d\rho^{2}-\eta^{2}\rho^{2} d\varphi^{2}- d z^{2}.
\end{equation}
where $\eta=1-4\lambda G/c^2$ is the deficit angle and $\lambda$ is the linear mass density of the cosmic string. In general, the deficit angle can assume $\eta>1$, which corresponds to an anti-conical spacetime with negative curvature. The geometry (\ref{line-element-1}) corresponds to a conical singularity described by the curvature tensor $R_{\rho,\varphi}^{\rho,\varphi}=\frac{1-\eta}{4\eta}\delta_{2}(\vec{r})$. The vierbein for this metric reads
\begin{equation}\label{vierbein}
    e^{a}_{~\mu}=
    \left(
      \begin{array}{cccc}
        1 & 0 & 0 & 0 \\
        0 & \cos\varphi & -\eta\rho\sin\varphi & 0 \\
        0 & \sin\varphi & \eta\rho\cos\varphi & 0 \\
        0 & 0 & 0 & 1 \\
      \end{array}
    \right)
\end{equation}
and
\begin{equation}\label{vierbein}
    e^{\mu}_{~a}=
    \left(
      \begin{array}{cccc}
        1 & 0 & 0 & 0 \\
        0 & \cos\varphi & \sin\varphi & 0 \\
        0 & -\frac{\sin\varphi}{\eta\rho} & \frac{\cos\varphi}{\eta\rho}& 0 \\
        0 & 0 & 0 & 1 \\
      \end{array}
    \right).
\end{equation}
The flat spacetime can be recovered for $\eta=1$. In the system of rectangular coordinates, the corresponding $\Omega$ matrix,
\begin{equation}\label{vierbein-c}
    \Omega^{a}_{~\mu}= (1-\eta)
    \left(
      \begin{array}{cccc}
        0 & 0 & 0 & 0 \\
        0 & -\sin^{2}\phi & \sin\phi\cos\phi & 0 \\
        0 & \sin\phi\cos\phi & -\cos^{2}\phi& 0 \\
        0 & 0 & 0 & 0 \\
      \end{array}
    \right)~.
\end{equation}
Inserting this into (\ref{effectiveA-1}) we can obtain the explicit expression of the effective vector potential, 
\begin{equation}\label{effectiveA-11}
\vec{K} 
= \frac{\chi }{q} \bigg(  \frac{qBx^{2}}{\rho} - \frac{l_{z}}{\rho} \bigg) \vec{e}_{\phi}~,
\end{equation}
where $\chi=1-\eta $. This correction is proportional to the angular momentum of the particle and the external magnetic field. Further, the topological defect affects only the physics on the azimuthal direction. This is due to the scaling of the azimuthal angle in the line element (\ref{line-element-1}). We will use the perturbation method to calculate the influences of this effective vector potential on the expectation values of the electric currents. In this approximation, the expectation values of the various currents are calculated by using the leading order wave-functions which are obtained by neglecting the $\chi$-dependent terms in the full Hamiltonian. The leading order results can be found in the standard textbooks, {\it e.g.}, in Ref. \cite{QM}.

The electric current can be obtained by using the Heisenberg equation. For the Hamiltonian (\ref{minic-ps-hamil}) we obtain
\begin{eqnarray}
J_{x} &=& q\rho_{c} \bigg( \frac{p_{x}}{m}  + \chi\frac{qBx^{2}y}{m\rho^{2}} - \chi\frac{y l_{z}+l_{z}y}{m\rho^{2}} \bigg)~,~ \\
J_{y} &=& q\rho_{c} \bigg( \frac{p_{y}}{m} - \frac{qB}{m}x  - \chi\frac{qBx^{3}}{m\rho^{2}} + \chi\frac{x l_{z}+l_{z}x}{m\rho^{2}} \bigg)  ~,~ \\
J_{z} &=& q\rho_{c} \frac{p_{z}}{m}  ~.
\end{eqnarray}
Then at the leading order the expectation values of these current operators are $\overline{J_{i}}=\langle n, \ell, k_{z} | J_{i} | n, \ell, k_{z} \rangle $ which have the following forms,
\begin{eqnarray}
\overline{J}_{x} &=&  q\rho_{c} \overline{ \bigg[ \frac{p_{x}}{m} + \frac{\chi}{m}\bigg( \frac{2qBx^{2}y}{\rho^{2}} - 2\Re \frac{yl_{z}}{\rho^{2}} \bigg) \bigg] }~,~  \\
\overline{J}_{y} &=&  q\rho_{c}\overline{\bigg[ \frac{p_{y}}{m} - \frac{qB}{m}x - \chi \bigg(  \frac{2qBx^{3}}{m\rho^{2}} - 2\Re \frac{xl_{z}}{m\rho^{2}} \bigg) \bigg]} ~,~ \\
\overline{J}_{z} &=& q\rho_{c} \frac{ \overline{p_{z}}}{m}~.
\end{eqnarray}
The expectation value of the operator $x^{2}y/\rho^{2}$ can be easily calculated by using the symmetry as follows.
\begin{equation}
\langle k_{y} | \frac{x^{2}y}{\rho^{2}} |  k_{y} \rangle
= \frac{1}{L_{y}} \int_{-a}^{a} d y \frac{x^{2}y}{\rho^{2}}
= - \frac{1}{L_{y}} \int_{-a}^{a} d y \frac{x^{2}y}{\rho^{2}}
=0
\end{equation}
where $a=L_y/2$, and $L_y$ is the length of the sample along the $\hat{y}$-direction. Thus, this term does not contribute the electric current along the $\hat{x}$-direction. The expectation value of the operator $y l_{z} /\rho^{2}$ is slightly complicated.  Integrating out the variable $y$ we obtain
\begin{equation}
\overline{\frac{y l_{z} }{\rho^{2}}}
=  \langle n |\bigg( \frac{2x}{L_{y}} \arctan (\frac{L_{y}}{2x}) - 1\bigg)p_{x} | n \rangle~,
\end{equation}
Noticing that $p_{x} | n \rangle$ is purely image\cite{QM}, then the expectation value of this operator is also pure image. Thus it does not also give any contribution. The expectation value of the operator $x^{3}/\rho^{2}$ is more complicated. Integrating our the variable $y$ we obtain
\begin{equation}
\overline{ \frac{x^{3}}{\rho^{2}} } 
= \langle n | \frac{2x^{2}}{L_{y}}\arctan( \frac{L_{y}}{2x} ) | n \rangle~.
\end{equation}
Due to the fact that the quantum fluctuation of the position operator is very small for harmonic oscillator, we can neglect these effects in the order of $\chi$. Then we obtain the approximated expectation value,
\begin{equation}
\overline{ \frac{x^{3}}{\rho^{2}} } 
\approx \frac{2 x_{c}^{2}}{ L_{y} }\arctan( \frac{L_{y}}{2x_{c}})~.
\end{equation}
Using the same approximation we can obtain the expectation value of the operator $xl_{z}/\rho^{2}$,
\begin{equation}
\overline{ \frac{x l_{z}}{\rho^{2}} }
= \langle n | \frac{\hbar k_{y} x^{2}}{\rho^{2}} | n \rangle
\approx  \hbar k_{y} \frac{2 x_{c}}{ L_{y} } \arctan( \frac{L_{y}}{ 2 x_{c}}) ~.
\end{equation}
Then collecting all these results we obtain the expectation values of the electric currents 
\begin{eqnarray}
\bar{J}_{x} &=& 0 ~, \\ 
\bar{J}_{y} &=& q\rho_{c} \frac{E}{B} \big( 1+ 2 \chi M( \kappa ) \big) ~, \\
\bar{J}_{z} &=& q\rho_{c} \frac{\hbar k_{z}}{m}
\end{eqnarray}
Here we define a dimensionless function 
\begin{equation}
M(\kappa)= \kappa \arctan( \frac{1}{ \kappa })~,~
\kappa = \frac{2x_{c}}{L_{y}}
\end{equation}
It is aas functions of the momentum in the $\hat{y}$-direction $k_{y}$, the external fields $E$ and $B$, center of the harmonic oscillator $x_{c}$ as well as the size of the sample $L_{y}$. In terms of the mass density of the topological defect we have $\chi=4\lambda G/c^2$. Then corrected Hall conductivity is
\begin{equation}
\tilde{\sigma}_{H}
=\sigma_{H} \bigg( 1+ \frac{8\lambda G}{c^2} M( \kappa ) \bigg)
\end{equation}

We study the Hall effect in the presence of topological dislocation.  The line element is
\begin{equation}\label{line-element-2}
     d s^{2}
    =c^2d t^{2}-d\rho^{2}-\rho^{2}d\varphi^{2}-(d z + \xi d\varphi)^{2}.
\end{equation}
where $\xi$ is the torsion of the topological dislocation\cite{Galtsov:1993,Lorenci:2002}. Topological dislocations are much more realistic line defects. They can modify the energy spectrum of electrons moving in a uniform magnetic field. Landau levels in the presence of dislocations have been investigated. The torsion can be identified with the surface density of the Burgers vector in the classical theory of elasticity. The vierbein in this case reads
\begin{equation}\label{vierbein}
    e^{a}_{~\mu}(x)=
    \left(
      \begin{array}{cccc}
        1 & 0 & 0 & 0 \\
        0 & \cos\varphi & -\rho\sin\varphi & 0 \\
        0 & \sin\varphi & \rho\cos\varphi & 0 \\
        0 & 0 & \xi & 1 \\
      \end{array}
    \right),
\end{equation}
and,
\begin{equation}\label{vierbein}
    e^{\mu}_{~a}(x)=
    \left(
      \begin{array}{cccc}
        1 & 0 & 0 & 0 \\
        0 & \cos\varphi & \sin\varphi & 0 \\
        0 & -\frac{\sin\varphi}{\rho} & \frac{\cos\varphi}{\rho}& 0 \\
        0 & \frac{\xi}{\rho}\sin\varphi & -\frac{\xi}{\rho}\cos\varphi & 1 \\
      \end{array}
    \right).
\end{equation}
The flat spacetime can be recovered for $\xi=0$.  In the reference frame of the rectangular coordinates, the corresponding $\Omega$ matrix is,
\begin{equation}\label{vierbein-c}
    \Omega^{a}_{~\mu}= 
    \left(
      \begin{array}{cccc}
        0 & 0 & 0 & 0 \\
        0 & 0 & 0 & 0 \\
        0 & 0 & 0 & 0 \\
        0 & -\frac{\xi}{\rho}\sin\phi  & \frac{\xi}{\rho}\cos\phi & 0 \\
      \end{array}
    \right).
\end{equation}
Inserting this into (\ref{effectiveA-1}) we can obtain the explicit expression of the effective vector potential,
\begin{equation}
\vec{K} =  \frac{\xi}{q} \bigg( \frac{l_{z}}{\rho^{2}} - \frac{qBx^{2}}{\rho^2} \bigg)\vec{e}_{z}.
\end{equation}
The correction again depends on the angular momentum and external magnetic field. Furthermore the components are the same as in last section. The only difference is that the correction is along the $\hat{z}$-direction rather than $\hat\phi$-direction. This is due to the coupling between $\phi$ and $z$ in the line element (\ref{line-element-2}). Again we use the perturbation theory to calculate the expectation values of the electric current operators in this case. Using the Heisenberg equation we obtain
\begin{eqnarray}
J_{x} &=& q\rho_{c} \bigg( \frac{p_{x}}{m} + \xi\frac{y}{\rho^{2}}\frac{p_{z}}{m} \bigg) ~,~ \\
J_{y} &=& q\rho_{c} \bigg( \frac{p_{y}}{m} - \frac{qB}{mc}x  - \xi\frac{x}{\rho^{2}}\frac{p_{z}}{m}\bigg)  ~,~ \\
J_{z} &=& \frac{q\rho_{c}}{m} \bigg( p_{z} - \xi \frac{l_{z}-qBx^{2}}{\rho^{2}}  \bigg)  ~.
\end{eqnarray}
Then at the leading order the expectation values of the current operators are
\begin{eqnarray}
\overline{J_{x}} &=& q\rho_{c} \frac{\xi \hbar k_{z}}{m}\langle n, k_{y} | \frac{y}{\rho^{2}} | n, k_{y} \rangle~,~  \\
\overline{J_{y}} &=& q\rho_{c} \bigg( \frac{\hbar k_{y}}{m} - \langle n, k_{y} | \frac{qB}{mc}x + \frac{\xi \hbar k_{z}}{m}\frac{x}{\rho^{2}} | n, k_{y} \rangle\bigg) , \\
\overline{J_{z}} &=& q\rho_{c} \bigg( \frac{\hbar k_{z}}{m} - \xi \langle n, k_{y} |\frac{l_{z}-qBx^{2}}{m\rho^{2}}  | n, k_{y} \rangle \bigg).
\end{eqnarray}
We have calculated the expectation value of the operator $\overline{y/\rho^{2}}$, it is zero at the leading order. By using the same approximation, we obtain the expectation values of the operators $x/\rho^{2}$,  $x^{2}/\rho^{2}$ and $l_{z}/\rho^{2}$,
\begin{eqnarray}
\overline{ \frac{x}{\rho^{2}} }
&\approx &  \frac{2}{L_{y}}\arctan( \frac{L_{y}}{ 2 x_{c}}) ~,
\\
\overline{ \frac{l_{z}}{\rho^{2}} }
&=& \overline{ \frac{\hbar k_{y} x}{\rho^{2}} }
\approx  \frac{2\hbar k_{y} }{L_{y}}\arctan( \frac{L_{y}}{ 2 x_{c}}) ~.
\end{eqnarray}
Then collecting all these results we obtain the expectation values of the electric current operators,
\begin{eqnarray}
\overline{J}_{x} &=& 0 ~,  \\
\overline{J}_{y} &=& - q\rho_{c} \frac{E}{B}\bigg( 1 - \frac{2\xi}{L_{y}} \frac{ \hbar k_{z} }{m} \frac{ B}{E} \frac{1}{\kappa} M(\kappa) \bigg) ~, \\
\overline{J}_{z} &=& q\rho_{c} \bigg( \frac{\hbar k_{z}}{m} + \frac{2\xi}{L_{y}} \frac{\hbar k_{y}}{m}  \frac{1}{\kappa}  M ( \kappa )  \bigg)~.
\end{eqnarray}
The corrected Hall conductance is 
\begin{equation}
\tilde{\sigma}_{H}
=\sigma_{H} \bigg( 1 - \frac{2\xi}{L_{y}} \frac{ \hbar k_{z} }{m} \frac{ B}{E} \frac{1}{\kappa} M(\kappa)  \bigg)~.
\end{equation}
Compared to the results in the case of topological defect, the corrected Hall conductance depends on the momentum in the $\hat{z}$-direction.

In summary, the influences of topological defect and dislocation on quantum Hall effects have been studied. For the topological defect with line element (\ref{line-element-1}), the vector potential  $\vec{A}$ receives corrections proportional to the linear density of the cosmic string, angular momentum $l_{z}$ of the particle and itself. These additional potentials affect only the dynamics along the azimuthal direction. By using the perturbation method we obtained the expectation values of the charge current $\overline{J}_{i}$ as well as the Hall conductance $\sigma_{H}$. The charge current $\overline{J}_{x}$ does not receive correction due to the harmonic motion along this direction, and $\overline{J}_{z}$ does not also influence due to the fact that the line element does not deform the motion of particle along this direction. However, the Hall current $\overline{J}_{y}$ receives correction proportional to the linear density of the cosmic string. The correction is always positive and depends on the expectation value of the coordinate $\overline{x}=x_{c}$ as well as the sample size $L_{y}$. The dependence is represented by a function $M(\kappa)=\kappa\arctan(1/\kappa), \kappa=2x_{c}/L_{y}$. For a light cosmic string with $\chi = 4\lambda G / c^2 \sim 10^{-9}$, it is very difficult to observe the effects. However for a cosmic string with  heavier mass density, the correction could be observable.

For the topological dislocation with line element (\ref{line-element-2}), the canonical momentum is deformed and receives a correction  proportional to the Ragga vector $\vec{\xi}$ and the angular momentum $l_{z}$ of the charge particle. The charge current $\overline{J}_{x}$ does not receive correction due to the harmonic motion along this direction. There is a correction along the $\hat{z}$-direction, this is a distinct feature from the last case. Even though $\overline{J}_{z}$ is affected, the effect will be negligible due to the smallness of the Raaga vector compared with the sample size $L_{y}$. However the Hall current $\overline{J}_{y}$ receives significant correction due to the enhancement factor $B/E$ which is of order $10^{8}$ typically. The correction also depends on the momentum along the $\hat{z}$ direction, and may be positive and negative depending on the expectation value of the coordinate $\overline{x}=x_{c}$, with $ - L_{x}/2 < x_{c} < L_{x}/2 $. For a Ragga vector of order $1 \rm{nm}$, the correction is of order $10^{-3}$ that is measurable.

\vskip 0.5cm \noindent\textbf{Acknowledgments}: K. M. is supported by the China Scholarship Council under Grant No. 201207010002, and the Hanjiang Scholar Project of Shaanxi University of Technology. J. H. W. is supported by the National Natural Science Foundation of China under Grant No. 11147181 and the Scientific Research Project in Shaanxi Province under Grant No. 2009K01-54 and Grant No. 12JK0960. H.-X. Y. is supported in part by the Startup Foundation of the University of Science and Technology of China and the Project of Knowledge Innovation Program (PKIP) of the Chinese Academy of Sciences.

\end{document}